\documentstyle[12pt, aaspp4, graphics]{article}

\begin{document}

\title{The Peak Brightness of SN1974G in NGC4414 and the Hubble Constant}
\author{Bradley E. Schaefer\altaffilmark{1,2}
\altaffiltext{1}{schaefer@grb2.physics.yale.edu}
\altaffiltext{2}{Visiting Astronomer,
Cerro Tololo Inter-American Observatory, operated by
AURA, Inc., under contract with the National Science Foundation.}
}
\affil{
Department of Physics, Yale University, PO Box 208121, New Haven CT 06520-8121
}

%This is single spacing
%\baselineskip 12pt
%This is double spacing
\baselineskip 24pt

\begin{abstract}

	The light curve of the Type Ia supernova SN1974G (in NGC4414) is 
important because the Hubble Space Telescope has measured the distance to
the host galaxy by means of Cepheid variables and thus the Hubble Constant 
can be derived.  Light curves from the secondary literature are inadequate 
since the majority of data is misreported, the majority of the published 
data is overlooked, and the majority of all data is unpublished, while 
comparison star sequences have offsets of over half a magnitude.  I have 
recovered and validated all data, remeasured the comparison stars, and 
performed light curve template fits.  I find the observed peak B and V
magnitudes to be $12.48 \pm 0.05$ and $12.30 \pm 0.05$, with a decline
rate of $\Delta m_{15} = 1.11 \pm 0.06$.  For $E(B-V) = 0.16 \pm 0.07$, 
the unabsorbed peak magnitudes are $B = 11.82 \pm 0.29$ and $V = 11.80 \pm
0.22$. With the distance modulus to NGC4414 as $\mu = 31.41 \pm 0.23$, I
find $H_{0} = 55 \pm 8 km \cdot s^{-1} \cdot Mpc^{-1}$.

\end{abstract}


\clearpage
\section{Introduction}

	Type Ia supernovae are perhaps the best standard candle (Branch \&
Tammann 1992 and references therein, Hamuy et al. 1996a) which can be used to 
measure the Hubble Constant ($H_{0}$).  However, to calibrate the peak
absolute magnitude relation of Hamuy et al. (1996a), both distances and peak 
magnitudes of historical supernovae are needed.  The Hubble Space Telescope has
recently measured the distance to several host galaxies containing Type Ia 
events (Saha et al. 1994, 1995, 1996ab; Sandage et al. 1996).  The key program 
on the distance scale has now observed NGC 4414, the host galaxy for SN1974G.  
Before this Cepheid distance can be of use, a reliable peak magnitude for the 
event must be known.

	How can the light curve of SN1974G be improved at this late date?  (1) 
I have made accurate measures of the brightnesses of the old comparison stars 
on the B and V magnitude systems, to allow modern reduction of the old 
observations.  Indeed, I find errors of up to 0.72 mag in the old literature. 
 (2) The majority of the observations are either unpublished in archives or 
recently published.  Indeed, even the majority of old published observations 
have been overlooked in the modern secondary literature.  (3) A variety of 
errors have appeared in the secondary literature, in particular, about half of 
the observations are assigned the wrong color including most of the 
pre-maximum and near-maximum data.  (4) Modern template fitting techniques 
can significantly improve the derived peak magnitudes, provide quantitative 
measures of the uncertainties, and supply the needed decline rate.

	This paper is the eighth in a series (Schaefer 1994, 1995abcd, 1996ab) 
where I provide modern peak brightnesses for Hubble Constant purposes.  

\section{Comparison Stars}

	SN1960F was discovered by W. Burgat (Burgat et al. 1974) on 1974 April
 20, roughly a dozen days before maximum.  The new star appeared in the 
outskirts of the Sc galaxy NGC 4414 at around thirteenth magnitude.  Spectra 
revealed a normal Type Ia event with a peak around JD2442170 (Ciatti \&
Rosino 1977 [C\&R], Patchett \& Wood 1976 [P\&W], and Iye et al. 1975).
Many photometric observations are available; in Burgat et al. (1974),
P\&W, C\&R, Howarth (1974), Burkholder (1995), and various IAU Circulars 
(numbers 2664, 2666, 2668, 2671, and 2678).  In addition, the AAVSO 
data base has provided 164 V magnitudes.

	Almost all the IAU Circular magnitudes are correctly reported in 
either Burgat et al. (1974), Howarth (1974, the BAA data), or in the AAVSO 
data.  The observations reported in Burgat et al. (1974) use the AAVSO 
sequence (Scovil 1974), while most of the visual data are in the AAVSO 
archives (the few exceptions were added to the AAVSO data).  The two visual 
estimates of Hopp (1974) could not be used since the sequence is not stated.

	The comparison stars for all utilized observations were explicitly 
identified by the observer.  A concordance of the various labels is presented 
in Table 1.  Previous experience with photometric data from the 1970's 
suggests that the comparison sequences are likely to have typical errors of a 
third of a magnitude (see also Figure 2 of Patat et al. 1997).  Such problems 
would cause important systematic errors in a Hubble Constant determination for 
SN1974G.  Fortunately, such errors can be corrected even today by measuring 
accurate comparison star magnitudes.  

	I have measured the Johnson B and V magnitudes for all comparison stars
 on four separate and independent nights.  CCD images of the comparison stars 
were obtained with the 2.4 m Hiltner telescope on Kitt Peak (February 1998) and
 the 0.9 m telescope on Cerro Tololo (June 1995).  All nights were definitely 
photometric.  The calibration from standards (Landolt 1992) was carried out 
independently for each night, with standard star observations being made 
immediately before and after the SN1974G comparison star images.  Standard 
stars were chosen to provide a wide range of colors, with B-V generally ranging
 from -0.2 to +1.4 mag.  For the Kitt Peak observations, twelve standard stars 
were observed multiple times between air masses of 1.30 and 2.10.  For the 
Cerro Tololo observations, the nightly number of standard star measures is 25 
over air mass ranging from 1.16 to 2.41, with about half around an air mass of 
2.0.  Each comparison star was measured from three to eight times, with the 
usual rms scatter in these values of 0.01 to 0.02 mag. 
 
	The individual CCD images were bias subtracted and flattened (with sky 
flats) by the normal procedures.  The APPHOT package in IRAF was used to 
perform aperture photometry on both standard and comparison stars.  The FWHM 
of the seeing disk was 1.5'' or better, so a photometry aperture of 10''
diameter was chosen.  All stars (other than one standard star which was 
excluded) had no significant contamination by stars or galaxies within this 
photometry aperture.  For the standard stars, the measured instrumental 
magnitudes were fit to a linear transformation equation with terms for the air 
mass and for the B-V color.  The fitted extinction coefficients agree well with 
previous measures for the sites both by myself and from the literature.  The 
color terms were in all cases small, with coefficients less than 0.05.  The rms
 scatter of the standard stars around the fitted transformation equation was 
typically from 0.01 to 0.02 mag.  The night-to-night agreement was better than 
0.01 mag in the B band, but was 0.02 mag in the V band.  The statistical 
uncertainties associated with the individual measures of the standard and 
comparison stars were always less than 0.01 mag.  For each image, the 
comparison star magnitudes were derived from the measured instrumental 
magnitude and that night's fitted transformation equation.  The true color of 
the  comparison stars were found after one iteration with convergence of 
better than 0.01 mag.  The final B and V magnitudes for each comparison star 
(see Table 1) are calculated as an average over all measures.  The overall 
reliability of my comparison star magnitudes is likely $\pm 0.02$ mag.

	With these accurate comparison star magnitudes, I have searched for 
systematic errors in the values adopted for each series of photometric 
measures of SN1974G.  I find that in all cases there is a simple relation 
between the adopted magnitudes and my modern values:

\begin{equation}
	m_{ph} = m_{ph}(Burkholder) \cdot 0.84 + 2.60,
\end{equation}
\begin{equation}
	B = B(P\&W) \cdot 0.97 + 0.99,	
\end{equation}
\begin{equation}
	B = B(C\&R),
\end{equation}
\begin{equation}
	V = V(C\&R) \cdot 1.07 - 1.13,
\end{equation}
\begin{equation}
	V = V(BAA) + 0.15
\end{equation}
\begin{equation}
	V = V(AAVSO) + 0.11.
\end{equation}

The observed rms scatters are 0.07, 0.08, 0.07, 0.07, 0.18, and 0.06 mag.  
Color terms were sought in each case, but in no case was a significant color 
term found. Tsvetkov (1986) measured some comparison stars photographically, 
but he was 0.09 and 0.06 mag in error on average with rms scatters of 0.09 and 
0.06 in B and V respectively.

	Typical magnitude errors for the comparison stars from the literature 
are 0.2 mag, with up to 0.72 mag error.  Equations 1-6 show that the 
deviations are complex, so any evaluation of the peak brightness of SN1974G 
based on uncorrected magnitudes will have comparable errors.

\section{Light Curve}

	With accurate comparison star magnitudes, the original observations 
can be reduced to provide the brightness of the supernova on a modern 
magnitude scale.  The details of this analysis will vary with the original 
observer's procedure.  

	C\&R report both B and V measures from the Asiago supernova
survey.  Patat et al. (1997) point out that the observations are close to the 
B and V magnitude systems.  They also point out that the measurements of the 
supernova brightness were made by interpolating between the brightness of the 
next brighter and next fainter comparison stars.  In practice, I correct for 
the changed comparison star magnitudes by constructing a plot of the
$V - V_{C\&R}$ versus $V_{C\&R}$ for the comparison stars with line segments 
connecting successively fainter stars, and apply the interpolated correction 
appropriate for $V_{C\&R}(SN)$ to the value.  So for example, if C\&R had
said that SN1974G has B = 14.45 (i.e., halfway between their magnitudes
for comparison stars F and G), then I would deduce that the actual
magnitude was 
B = 14.48 (i.e., halfway between F and G).  This procedure is identical with 
that recommended by Patat et al. (1997) in their equation 3.

	The P\&W magnitudes and comparison star sequence are both
explicitly stated to be Johnson B.  Their comparison star sequence was 
established with iris diaphragm photometry and a quadratic calibration curve 
(magnitude versus iris reading) was fitted from all comparison stars for each 
plate.  The magnitude of the supernova was then determined from the iris 
reading for the supernova and the calibration curve.  For errors in the 
comparison star magnitudes that follow equation 2, the calibration curve 
and the supernova magnitude can be corrected by an application of equation 2.

	The AAVSO and BAA observations are from many visual observers.  
Multiple observations on the same night (or within a few days for late in the 
light curve) have been combined to form normal points.  The visual estimates 
taken before the AAVSO sequence was established (1974 April 27) are not used 
because their comparison stars are unknown.  Visual observers generally will 
interpolate the brightness between the next fainter and brighter comparison 
stars, so it is best to use a procedure similar to that used for the C\&R
analysis (i.e., like equation 3 of Patat et al. 1997).  The visual 
observations have a color sensitivity very close to Johnson V (Schaefer 1996a).

	The observations of Burgat et al. (1974) were on panchromatic film 
with the AAVSO visual sequence, and hence are closest to the V magnitude 
system.  The color term for the panchromatic plates is unknown and could be 
very large (Schaefer 1996a).  I will assume that the magnitudes were extracted 
from the plates by visual comparison with the next brighter and fainter stars, 
so I used the same analysis procedure as for the AAVSO data.

	Burkholder (1995) reports microdensitometry of five plates, and she 
presents instrumental magnitudes for both comparison stars and the supernova 
in the photographic magnitude system.  The relation $m_{ph} = 
B + 0.18(B - V) - 0.29$ is valid for stars and was used to convert the 
comparison star magnitudes to the photographic system.  My reanalysis followed 
Burkholder's Table 4, except with the substitution of these modern comparison 
star magnitudes.  The resulting photographic magnitudes for the supernova were 
then converted to B magnitudes with the above relation (which is known to be 
valid for supernovae) for B-V colors iteratively determined from all 
SN1974G observations.

	A concern with all supernova photometry is the contamination from the 
background galaxy light.  Fortunately, SN1974G appeared on the outside edge of 
NGC4414 so the contamination is minimal.  This can be explicitly seen for the 
plates of Burgat, P\&W, C\&R, and Howarth (see their Figure 1, Plate II,
Figure 8, and Figure 2 respectively), in that the SN image is at a location of 
no significant recorded flux from NGC4414.  To be quantitative, I have measured
 the surface brightness of NGC4414 at the location of the supernova, with 
surface brightnesses of $B \sim 22.2$ and $V \sim 21.6$ magnitudes per
square arc-second.  This is to be compared with typical dark zenith sky
brightnesses of $B \sim 22.5$ and $V \sim 21.5$ magnitudes per square
arc-second, or with sky brightnesses of $B \sim 19.2$ and $V \sim 18.7$
magnitudes per square arc-second on nights (JD around 2442171, 2442198, and 
2442228) with the near full Moon $\sim 40^{o}$ away.  Within the typical
aperture for  iris diaphragm photometry (close to the star size with
radius of $\sim 3''$ for the late-time plates of P\&W and C\&R; Schaefer
1982), the galaxy contributes only the brightness of a 18.6 and a 17.9 mag 
star (in B and in V).  For iris diaphragm photometry, the measurement is 
essentially that of the image radius to some isophotal level, while this 
radius will increase slightly with light from the host galaxy.  For the 
P\&W images, the plate limiting magnitude ($B \sim 16$) corresponds to a
toe radius with an isophotal level of $B \sim 17$ magnitudes per square
arc-second, such that the galactic background changes the brightness at
the toe by 0.008 magnitudes per square arcsecond, and hence an error in the 
deduced SN magnitude of 0.008 mag.  In all cases, this 
added light is negligible even for late times in the light curve of SN1974G.

	Following Schaefer (1994), I adopt a one-sigma uncertainty of 0.15 mag 
for magnitudes from photographs.  However, the panchromatic photographs will 
be assigned an uncertainty of 0.5 mag to allow for the expected large and 
unknown color terms.  Individual visual observations are assigned an 
uncertainty of 0.3 mag (0.6 mag if the magnitude is reported with a
``:''), 
which is then reduced by the square root of the number of observations used 
to form the normal point.

	Tables 2 and 3 summarize the B and V light curves for SN1974G, while 
Figure 1 plots the light curves.  The magnitudes from the AAVSO and the BAA 
are indicated in the `Source' column of Table 3 by either an `A' or a `B'
followed in parentheses by the number of contributing observations.

\section{Peak Magnitude}

	This light curve can yield an accurate peak magnitude and decline rate 
by the normal template fitting techniques.  I have used the six templates 
presented in Hamuy et al. (1996b).  Specific procedures that I have used are 
described in Schaefer (1996b).

	The best combined fit to all the B and V data yields a B peak on 
$JD2442168.5 \pm 0.5$ for the SN1992al template.  The peak brightness is
$B = 12.48 \pm 0.05$ and $V = 12.30 \pm 0.05$.  The decline rate has
$\Delta m_{15} = 1.11 \pm 0.06$.  The resultant reduced $\chi^{2}$ is 1.60
(37 B points and 51 V points), with all the excess above unity coming from 
four V brightnesses long after peak.

	The robustness of this result can be tested by varying the choices 
used in the above analysis:  (1) The B and V data can be fit separately.  The 
best B fit is for the SN1992al template with a peak on $JD2442168.0 \pm 
0.5$ at $B = 12.45 \pm 0.05$.  The best V fit is for the SN1992al template with
 a peak on $JD2442169.0 \pm 0.5$ at $V = 12.32 \pm 0.05$.  (2) The 21 B and 27 
V data points within 20 days of peak can be fit.  The best combined fit is for 
the SN1992al template with a peak on JD2442169.0 at $B = 12.56 \pm 0.08$
and $V = 12.27 \pm 0.07$  (with a reduced $\chi^{2}$ of 1.22).   (3) The
11 B and 11 V observations of C\&R can be fit by themselves.  The best
combined fit is for the SN1992al template (with a reduced  $\chi^{2}$ of
0.82) for a peak on $JD2442169.0 \pm 0.6$ at $B = 12.42 \pm 0.10$ and
$V = 12.18 \pm 0.09$.  (4) If equations 1-6 are used for all data
sets, then the best combined fit is for the SN1992al template (with a
reduced $\chi^{2}$ of 1.58) for a peak on $JD2442168.5 \pm 0.5$ at
$B = 12.48 \pm 0.05$ and $V = 12.29 \pm 0.05$.  (5) A direct average of 
the 7 B and 5 V points within 4 days of peak, after corrections for the 
SN1992al template for a peak on JD2442168.5, is $B = 12.53 \pm 0.06$ and
$V = 12.30 \pm 0.07$.  A separate linear fit to the B data between 10-20 days 
after peak gives a fitted magnitude at 15 days after peak of
$B = 13.66 \pm 0.06$, for a value of $\Delta m_{15}$ equal to $1.13 \pm
0.09$ mag.  The scatter of these five alternative fits is consistent with
the quoted error bars for the preferred fit, so I will adopt the best fit 
from the 
previous paragraph ($JD2442168.5 \pm 0.5$, $B = 12.48 \pm 0.05$, $V = 
12.30 \pm 0.05$, and $\Delta m_{15} = 1.11 \pm 0.06$).

	The peak magnitudes should be corrected for the extinction.  The 
extinction from our Milky Way should be near zero (Leibundgut et al. 1991).  
SN1974G appears far outside NGC4414, so the host galaxy extinction is expected 
to be small.  The E(B-V) value can be quantitatively estimated by three means: 
 (1) The various spectra show no indication of sodium absorption lines, with 
the most restrictive being the C\&R spectra for which the equivalent width
is less than roughly 0.25.  The correlation between theequivalent width and 
E(B-V) is loose (Ho \& Filippenko 1995; Richmond et al. 1994; Barbon et
al. 1990), but the C\&R limit implies E(B-V) can plausibly be as high as
$\sim 0.2$ mag.  (2) The B-V color at peak is $0.18 \pm 0.07$, which
implies noticeable extinction.  With the unreddened color at peak of
$B - V = 0.00 \pm 0.04$ (Schaefer 1995c), the value of 
E(B-V) is $0.18 \pm 0.08$.  (3) The third means of estimating the
reddening is to use the result from Riess, Press, \& Kirshner (1996) that
all Type Ia events have the same intrinsic color at 45 days after peak, with 
$(B-V)_{0} = 1.00 \pm 0.10$.  By fitting the SN1974G light curve from
$JD2442200-2442220$, I find $B - V = 1.11 \pm 0.11$, for a resulting $E(B-
V) = 0.11 \pm 0.15$.  From these three methods, I conclude that
$E(B-V) = 0.16 \pm 0.07$.

	For $A_{V}/E_{B-V} = 3.1$, I find $A_{B} = 0.66 \pm 0.29$ and
$A_{V} = 0.50 \pm 0.22$. Thus, my final peak brightnesses for SN1974G are
$B = 11.82 \pm 0.29$ and $V = 11.80 \pm 0.22$.

\section{Hubble Constant}

	In the past several years, I have been calibrating the peak magnitudes 
of nearby Type Ia supernovae for which Cepheid distances have been measured 
with the Hubble Space Telescope.  Just as with SN1974G, the old light curves 
often require significant improvements before they can be reasonably used for 
any analysis of the Hubble constant.  It must be realized, that without these 
reliable light curves, the large number of orbits used by the Hubble Space 
Telescope have only poor utility for calibrating the supernovae.

	Hamuy et al. (1996a) has resolved the debate over the dependence of 
Type Ia luminosity on the decline rate.  Events with varying decline rate can 
be corrected to a standard decline rate with

\begin{equation}
	M_{o}^{*} = m_{max} - A - \mu - b^{*}[\Delta m_{15}-1.1]
\end{equation}

Here, $M_{o}^{*}$ is the peak absolute magnitude corrected for reddening
and decline rate, $m_{max}$ is the apparent peak magnitude, A is the 
extinction, $\mu$ is the distance modulus to the host galaxy, and
$\Delta m_{15}$ is the decline rate as measured by the magnitudes below
peak in the B filter the light curve is 15 days after peak.  The
parameter `b' is the slope of the decline rate relation; $b = 0.78 \pm 
0.18$ in B and $b = 0.71 \pm 0.14$ in V (Hamuy et al. 1996a).  Hamuy 
et al. (1996c) presents a Hubble diagram for distant supernovae, from which 
they derive $M_{o}^{*}$ values for some assumed fiducial Hubble constant.
They find (for $H_{0} = 50 km \cdot s^{-1} \cdot Mpc^{-1}$) that
$M_{o,50}^{*}$ values
are $-19.83 \pm 0.05$ in the B and $-19.84 \pm 0.04$ in the V.

	SN1974G has $M_{o}^{*}$ values of $-19.60 \pm 0.37$ in the B and
$-19.62 \pm 0.31$ in the V. When applied to the distant Hubble diagram, the 
derived Hubble Constant will be $100.2(M_{o}^{*} - M_{o,50}^{*}) \cdot 50
km \cdot s^{-1} \cdot Mpc^{-1}$. This leads to $H_{0}$ values of $55 \pm
10$ and $55 \pm 8 km \cdot s^{-1} \cdot Mpc^{-1}$ in B and V respectively.
These two values are
not independent since the B-V color was used to derive the reddening.  So
the conclusion is that SN1974G yeilds a calibration of the peak brightness for 
Type Ia supernovae that corresponds to a Hubble constant of $55 \pm 8 km
\cdot s^{-1} \cdot Mpc^{-1}$.

I thank the many members of the American Association of Variable Star 
Observers, as well as Jonathan Kemp for help with the MDM observations.

\begin{table}
\begin{center}
\begin{tabular}{|c|c|c|c|c|c|}
\hline
P\&W	& C\&R 	& BAA	& AAVSO	& B     & V    \\
\hline
1	&	&	&	&10.73	&10.05 \\
2	&	&	&	&11.53	&10.80 \\
3	&	&	&	&12.11	&11.21 \\
4	& C	&12.1	&123	&13.12	&12.44 \\
5	&	&12.5	&122	&13.32	&12.34 \\
6	&	&13.7	&138	&14.50	&13.93 \\
7	& I	&14.3	&143	&14.91	&14.40 \\
8	&	&13.0	&131	&14.13	&13.30 \\
9	&	&	&	&15.35	&14.56 \\
11	&	&	&	&15.85	&15.17 \\
	&B	&	&	&12.59	&11.48 \\
	&D	&	&	&13.65	&12.65 \\
	&E	&	&	&13.70	&12.98 \\
	&F	&	&	&14.35	&13.59 \\
	&G	&	&	&14.61	&13.99 \\
	&L	&14.8	&148	&15.90	&14.89 \\
	&N	&	&151	&15.85	&15.11 \\
	&P	&	&	&16.52	&15.40 \\
	&	&	&148	&16.34	&14.87 \\
\hline
\end{tabular}
\caption{Comparison star magnitudes.}
\end{center}
\end{table}

\begin{table}
\begin{center}
\begin{tabular}{|c|c|c|c|c|c|}
\hline
JD- 	&	&	&JD- & & \\
2442000 &B & Source & 2442000 &B &Source \\
\hline
164.44 &$12.65 \pm 0.15$ &C\&R &187.58 &$14.07 \pm 0.15$ &Burk \\
164.50 &$12.72 \pm 0.15$ &P\&W &188.61 &$13.91 \pm 0.15$ &Burk \\
166.50 &$12.57 \pm 0.15$ &P\&W &189.50 &$14.38 \pm 0.15$ &P\&W \\
167.50 &$12.51 \pm 0.15$ &P\&W &190.38 &$14.17 \pm 0.15$ &C\&R \\
168.50 &$12.56 \pm 0.15$ &P\&W &192.50 &$14.66 \pm 0.15$ &P\&W \\
169.50 &$12.46 \pm 0.15$ &P\&W &193.44 &$14.41 \pm 0.15$ &C\&R \\
169.50 &$12.53 \pm 0.15$ &P\&W &194.40 &$14.76 \pm 0.15$ &C\&R \\
176.50 &$12.96 \pm 0.15$ &P\&W &194.52 &$14.71 \pm 0.15$ &C\&R \\
177.36 &$12.71 \pm 0.15$ &C\&R &197.50 &$15.07 \pm 0.15$ &P\&W \\
177.58 &$13.26 \pm 0.15$ &Burk &203.50 &$15.41 \pm 0.15$ &P\&W \\
180.45 &$13.23 \pm 0.15$ &C\&R &205.50 &$15.50 \pm 0.15$ &P\&W \\
180.50 &$13.19 \pm 0.15$ &P\&W &210.37 &$15.46 \pm 0.15$ &C\&R \\
181.50 &$13.48 \pm 0.15$ &P\&W &212.39 &$15.84 \pm 0.15$ &C\&R \\
181.58 &$13.64 \pm 0.15$ &Burk &213.50 &$15.47 \pm 0.15$ &P\&W \\
182.50 &$13.59 \pm 0.15$ &P\&W &218.50 &$15.76 \pm 0.15$ &P\&W \\
182.57 &$13.77 \pm 0.15$ &Burk &219.50 &$15.62 \pm 0.15$ &P\&W \\
183.43 &$13.54 \pm 0.15$ &C\&R &221.50 &$15.56 \pm 0.15$ &P\&W \\
185.50 &$13.94 \pm 0.15$ &P\&W &245.38 &$16.02 \pm 0.15$ &C\&R \\
186.50 &$14.10 \pm 0.15$ &P\&W &       &		 & \\
\hline
\end{tabular}
\caption{B Light curve for SN1974G.}
\end{center}
\end{table}

\begin{table}
\begin{center}
\begin{tabular}{|c|c|c|c|c|c|}
\hline
JD-	& &		&JD- & & \\
2442000	&V &Source &2442000 &V &Source \\
\hline
157.50	&$13.79 \pm 0.50$ &Burgat &188.70 &$13.40 \pm 0.09$ &A(11) \\
160.40	&$12.69 \pm 0.50$ &Burgat &190.39 &$12.82 \pm 0.15$ &C\&R \\
160.50	&$12.79 \pm 0.50$ &Burgat &191.60 &$13.76 \pm 0.11$ &A(7) \\
164.46	&$12.22 \pm 0.15$ &C\&R	&193.40	&$13.72 \pm 0.09$ &B(11) \\
166.50	&$12.40 \pm 0.17$ &B(3)	&193.46	&$13.35 \pm 0.15$ &C\&R \\
166.60	&$12.48 \pm 0.09$ &A(12) &193.50 &$12.99 \pm 0.50$ &Burgat \\
169.60	&$12.36 \pm 0.11$ &A(7)	&193.50	&$13.79 \pm 0.50$ &Burgat \\
172.60	&$12.32 \pm 0.11$ &A(8)	&193.50	&$13.87 \pm 0.50$ &Burgat \\
174.50	&$12.37 \pm 0.21$ &B(2)	&194.41	&$13.47 \pm 0.15$ &C\&R \\
175.60	&$12.47 \pm 0.09$ &A(11)&194.53	&$13.53 \pm 0.15$ &C\&R \\
177.38	&$12.43 \pm 0.15$ &C\&R	&194.60	&$13.58 \pm 0.07$ &A(17) \\
177.40	&$12.54 \pm 0.50$ &Burgat &197.40 &$14.00 \pm 0.21$ &A(2) \\
178.40	&$12.36 \pm 0.13$ &B(5)	&198.40	&$14.14 \pm 0.27$ &B(5) \\
178.60	&$12.50 \pm 0.09$ &A(12)&203.40	&$14.45 \pm 0.19$ &B(4) \\
180.46	&$12.54 \pm 0.15$ &C\&R	&206.60	&$14.20 \pm 0.11$ &A(7) \\
181.60	&$12.69 \pm 0.07$ &A(19) &210.38 &$14.18 \pm 0.15$ &C\&R \\
182.40	&$12.82 \pm 0.08$ &B(14) &210.60 &$14.58 \pm 0.17$ &A(3) \\
183.40	&$13.29 \pm 0.50$ &Burgat &211.50 &$14.75 \pm 0.17$ &B(3) \\
183.40	&$12.82 \pm 0.50$ &Burgat &212.41 &$14.22 \pm 0.15$ &C\&R \\
183.42	&$12.65 \pm 0.15$ &C\&R	&213.40	&$14.51 \pm 0.15$ &A(4) \\
184.50	&$12.93 \pm 0.08$ &A(13) &214.50 &$14.29 \pm 0.50$ &Burgat \\
186.40	&$13.09 \pm 0.50$ &Burgat &217.60 &$14.53 \pm 0.15$ &A(4) \\
186.40	&$12.85 \pm 0.09$ &B(10) &218.40 &$14.69 \pm 0.50$ &Burgat \\
186.50	&$13.03 \pm 0.10$ &A(9)	&224.60	&$14.85 \pm 0.15$ &A(4) \\
187.60	&$13.16 \pm 0.11$ &A(8)	&245.41	&$15.40 \pm 0.15$ &C\&R \\
188.40	&$12.86 \pm 0.13$ &B(5)   &	&		& \\
\hline
\end{tabular}
\caption{V Light curve for SN1974G.}
\end{center}
\end{table}

\begin{figure}
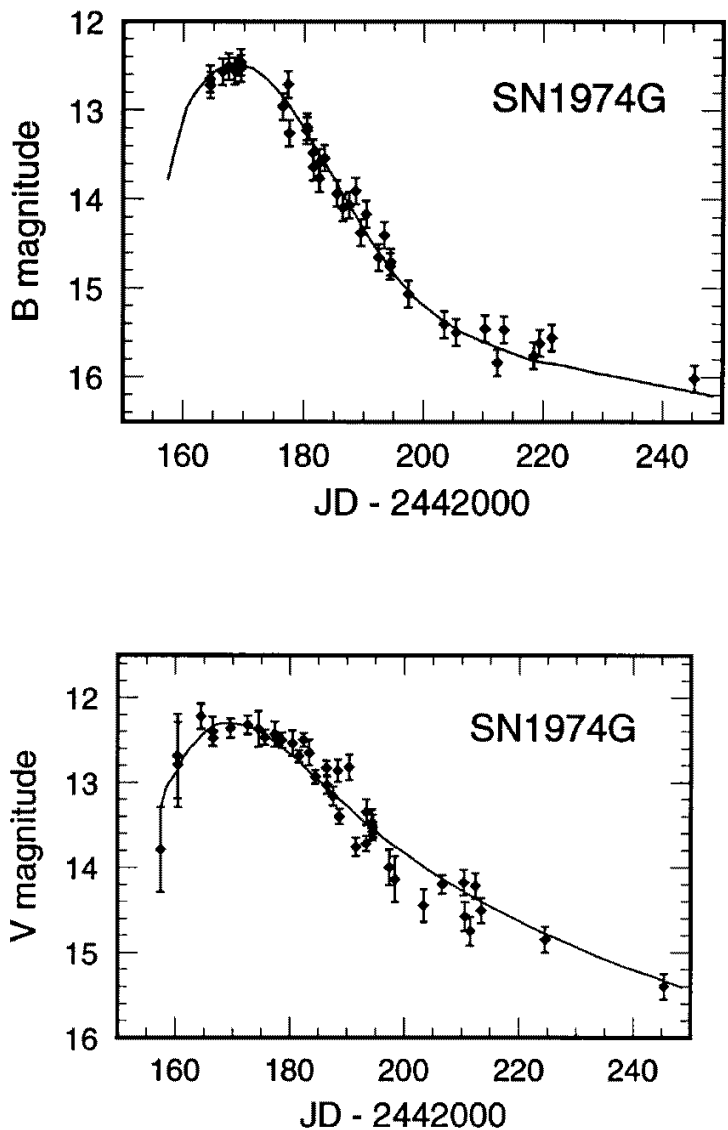

\begin{center}
\includegraphics{Bplot.ps}
\includegraphics{Vplot.ps}
\end{center}
\caption{Light curve for SN1974G. These B and V light curves for SN1974G
is composed of observations from many observers yet the scatter is
reasonably small, except for a few late V observations.  (The 
panchromatic points after peak are not displayed due to their very large 
uncertainty.) Template fits (smooth curves) show a peaks of $12.48 \pm 
0.05$ and $12.30 \pm 0.05$ in the B and V bands.  The B-V color at peak 
implies there is some small but significant extinction.  The good coverage in 
this light curve allow for a well defined decline rate ($\Delta m_{15} =
1.11
\pm 0.06$) and date of peak ($JD2442168.5 \pm 0.5$).
}
\end{figure}

\end{document}